\documentclass[aps,prd,twocolumn,showpacs,showkeys,superscriptaddress,footinbib]{revtex4-1}
\usepackage{amsmath}
\usepackage{bm}

\newcommand\vek[1]{\bm{#1}}
\newcommand\he[1]{#1^{\dagger}}
\newcommand\gr[1]{\mathrm{#1}}
\newcommand{\nbs}{n_{\mathrm{BS}}}
\newcommand{\nng}{n_{\mathrm{NG}}}
\newcommand{\dn}{\Delta n}
\DeclareMathOperator{\rank}{rank}
\newcommand{\bra}[1]{\langle{#1}|}
\newcommand{\ket}[1]{|{#1}\rangle}
\newcommand{\braket}[2]{\langle{#1}|{#2}\rangle}
\newcommand{\La}{\mathcal{L}}
\newcommand{\abs}[1]{|#1|}
\newcommand{\op}[1]{\hat{#1}}
\newcommand{\M}{\mathcal M}
\newcommand{\I}{\mathcal I}
\newcommand{\U}{\op{\mathcal U}}

\begin{document}

\title{On the number of Nambu--Goldstone bosons and its relation to charge densities}

\author{Haruki Watanabe}
\email{hwatanabe@berkeley.edu}
\affiliation{Department of Physics, University of Tokyo, Hongo, Tokyo 113-0033, Japan}
\affiliation{Department of Physics, University of California, Berkeley, California 94720, USA}

\author{Tom\'{a}\v{s} Brauner}
\email{tbrauner@physik.uni-bielefeld.de}
\affiliation{Faculty of Physics, University of Bielefeld, 33615 Bielefeld, Germany} 
\affiliation{Department of Theoretical Physics, Nuclear Physics Institute ASCR, 
25068 \v Re\v z, Czech Republic}

\begin{abstract}
The low-energy physics of systems with spontaneous symmetry breaking is governed by the associated
Nambu--Goldstone (NG) bosons. While NG bosons in Lorentz-invariant systems are well understood, the
precise characterization of their number and dispersion relations in a general quantum many-body
system is still an open problem. An inequality relating the number of NG bosons and their dispersion
relations to the number of broken symmetry generators was found by Nielsen and Chadha. In this
paper, we give a presumably first example of a system in which the Nielsen--Chadha inequality is
actually not saturated. We suggest that the number of NG bosons is exactly equal to the number of
broken generators minus the number of pairs of broken generators whose commutator has a nonzero
vacuum expectation value. This naturally leads us to a proposal for a different classification of
NG bosons.
\end{abstract}

\maketitle


\section{Introduction and summary}

Spontaneous symmetry breaking (SSB) is a ubiquitous phenomenon in Nature. In quantum many-body theory,
its significance is further boosted by the fact that it provides us with a rare example of a general
exact result: the Goldstone theorem~\cite{Goldstone:1961eq,Goldstone:1962es}. The low-energy physics
of systems exhibiting SSB is governed by the associated soft excitations, the Nambu--Goldstone (NG)
bosons (see Ref.~\cite{Guralnik:1968gu} for a comprehensive early review). While NG bosons in
Lorentz-invariant systems are well understood, the precise characterization of their number and
dispersion relations in a general quantum many-body system is still an open problem.

The systematic study of NG bosons in Lorentz-noninvariant systems was initiated by Nielsen and
Chadha in their seminal work~\cite{Nielsen:1975hm}. They showed that under certain technical
assumptions, including rotational invariance and the requirement of absence of long-range interactions, the
energy of the NG boson is proportional to an integer power of momentum in the long-wavelength
limit, $\varepsilon({\vek k})\propto\abs{\vek k}^n$. The NG boson is then classified as type-I if $n$ is
odd, and as type-II if $n$ is even. Nielsen and Chadha proved that \emph{the number of type-I NG
bosons plus twice the number of type-II NG bosons is greater than or equal to the number of broken
symmetry generators}.

In Lorentz-invariant systems, the Nielsen--Chadha (NC) inequality is trivial since it is well known that
the number of NG bosons, all being naturally type-I, equals the number of broken generators. In Lorentz-noninvariant 
systems, the NC inequality allows the number of NG bosons to be smaller than the
number of broken generators provided some of them have nonlinear dispersion relation. A renown
example is the ferromagnet where two generators broken by the spontaneous magnetization give rise to
only one NG boson---the magnon---having a quadratic dispersion relation at low momentum.

It is perhaps worth emphasizing that the NC counting rule is formulated as an inequality and that it does not constrain in any way the power of momentum appearing in the dispersion relation. In fact, a quick sociological survey reveals that this generality of the Nielsen--Chadha theorem is not very well appreciated in publications citing the original work~\cite{Nielsen:1975hm}. The reason for this probably is that in the concrete systems studied in literature, type-I and type-II NG bosons always have linear and quadratic dispersion relations, respectively, and the NC inequality is always saturated. In Appendix~\ref{app:inequality} of this paper, \emph{we point out a rather trivial class of theories where NG bosons with energy proportional to an in principle arbitrarily high power of momentum can appear, and the NC inequality is not necessarily saturated}~\footnote{The power of momentum in the dispersion relation is bounded from above by a value depending on the space dimension $d$ though. Generalizing naively the argument of the classic paper by Coleman~\cite{Coleman:1973ci} to Lorentz-noninvariant systems, the correlation function of the NG field would diverge at large distance if the power of momentum exceeded this limit.}. This ultimately leads us to the proposal for a different classification of NG bosons, and the evidence that this provides and exact \emph{equality} for their count.

Our work is based on the dependence of the number of NG bosons and their dispersion relations on
expectation values of conserved charges, which was first investigated at the general level by
Leutwyler~\cite{Leutwyler:1993gf} and Sch\"afer~\emph{et al.}~\cite{Schafer:2001bq} (see also
Ref.~\cite{Brauner:2010wm} for a recent review). Leutwyler~\cite{Leutwyler:1993gf} used effective
field theory to argue that, in the absence of quantum anomalies, nonzero density of a non-Abelian
conserved charge leads to the appearance of NG bosons with a quadratic dispersion relation. This
connection was further elaborated in Ref.~\cite{Brauner:2005di} where en exact equality (saturating
the NC inequality) was established for the class of linear sigma models with chemical potential at
tree level. Every pair of broken generators whose commutator has a nonzero expectation value was
found to give rise to one type-II NG boson with a quadratic dispersion relation. On the other hand,
Sch\"afer~\emph{et al.}~\cite{Schafer:2001bq} related charge densities to the number of NG bosons by
showing that this number equals the number of broken generators provided the expectation values of
commutators of all pairs of broken generators vanish.

Our present proposal can in a sense be understood as a synthesis of these previous works. We argue
that given the three characteristics of SSB, that is, the number of NG bosons, their dispersion
relations, and the expectation values of charge densities, an exact equality can be obtained by
focusing on the relation between the number of NG bosons and the charge densities. Thus, our work
generalizes the theorem of Sch\"afer~\emph{et al.}~\cite{Schafer:2001bq}. This is in contrast to
Nielsen and Chadha~\cite{Nielsen:1975hm} as well as Leutwyler~\cite{Leutwyler:1993gf} who emphasized
the role of the NG boson dispersion relations.

In order that the generality of our proposal is not obscured by the technical details, we formulate it here. Let $Q_a$ be the set of (spontaneously broken) conserved charges of the theory. \emph{The number of NG bosons $\nng$ is related to the number of broken symmetry generators $\nbs$ by the equality
\begin{equation}
\nbs-\nng=\frac12\rank\rho,
\label{conjecture}
\end{equation}
where the matrix of commutators $\rho$ is defined by
$i\rho_{ab}\equiv\lim_{\Omega\to\infty}\frac1\Omega\bra0[Q_a,Q_b]\ket0$} and $\Omega$ is the space volume. Note that $\rank\rho$ is always even since $\rho$ is real and antisymmetric. Unfortunately, we were not able to prove Eq.~\eqref{conjecture} in the full generality so it remains a conjecture at the moment. However, we can prove a one-side inequality and provide evidence that the opposite inequality holds as well. We should also emphasize that our argument applies to continuum field theories  as well as to models defined on a (space) lattice.

The paper is organized as follows. In Section~\ref{sec:generalSSB} we make some general remarks on
SSB, basically setting up the stage for the discussion of our main result. We also introduce a
class of ``uniform'' symmetries to which the standard derivation of the Goldstone theorem applies.
In Section~\ref{sec:proof} we give a partial proof of the conjecture \eqref{conjecture} and provide
evidence for the missing part of the proof. Finally, in Section~\ref{sec:conclusions} we conclude
and afford some speculations. Some technical details as well as specific examples that do not
pertain to the general argument, are relegated to the appendices.


\section{General remarks on SSB and the Goldstone theorem}
\label{sec:generalSSB}

In this section we will briefly review the notion of SSB and the Goldstone theorem. Although we will
not repeat in detail its proof, we would like to make a number of comments explaining under what
conditions and technical assumptions this proof applies. This may at times look like purely academic
pedantry, yet the example of a sharp inequality in the NC counting rule presented in
Appendix~\ref{app:inequality} shows that one should be prepared for the unexpected.

Let us first, following Nielsen and Chadha~\cite{Nielsen:1975hm}, define what we mean by the
number of broken generators $\nbs$. We demand that there is a set of conserved charges $Q_a$ and a
set of (quasi-)local fields $\Phi_a(x)$, or, on a \emph{space} lattice, operators $\Phi_a(t,\vek x_i)$, where $a=1,\dotsc,\nbs$, 
such that the matrix 
\begin{equation}
M_{ab}\equiv\bra0[Q_a,\Phi_b(0)]\ket0
\label{M}
\end{equation}
is nonsingular (i.e.~has a nonzero determinant). Moreover, we assume that the conserved charges can be
expressed as a spatial integral or sum of local charge densities,
\begin{equation}
\begin{split}
Q_a&=\int d^d\vek x\,j^0_a(t,\vek x),\quad\text{or}\\
Q_a&=\sum_{\vek x_i\in\text{lattice}}\rho_a(t,\vek x_i).
\end{split}
\label{Qdef}
\end{equation}
Strictly speaking, the broken charge operators are only well defined in a finite volume.
Nevertheless, in the formal derivations they only appear in commutators which have a well-defined
infinite-volume limit~\cite{Guralnik:1968gu}.


\subsection{Uniform symmetries}
\label{sec:uniform}

The proof of the Goldstone theorem essentially proceeds by finding the spectral representation of
the commutator \eqref{M}. This heavily relies on the integral representation \eqref{Qdef} and the
following translational property of the charge densities,
\begin{equation}
\begin{split}
j^0_a(x)&=e^{iP\cdot x}j^0_a(0)e^{-iP\cdot x},\quad\text{or}\\
\rho_a(t,\vek x_i)&=e^{iHt}\he T_{\vek x_i}\rho_a(0)T_{\vek x_i}e^{-iHt},
\end{split}
\label{translation}
\end{equation} 
where $P^\mu$ is the four-momentum operator, $H$ the Hamiltonian, and $T_{\vek x_i}$ is the operator of a finite (lattice) 
translation by $\vek x_i$. Even though this property is usually taken for granted, it only holds provided that the
charge density operator, as constructed in terms of the local operators of the theory, does not depend
\emph{explicitly} on the coordinate. We will call symmetries whose charge densities satisfy this condition
\emph{uniform}.

As an example, consider the simplest field theory exhibiting SSB and a NG boson: the free massless
relativistic scalar field theory, defined by the Lagrangian
\begin{equation}
\La=\frac12\partial^\mu\phi\partial_\mu\phi.
\end{equation}
The action of this theory is invariant under the set of global transformations
$\phi(x)\to\phi(x)+\theta(x)$ with $\theta(x)=a+b_\alpha x^\alpha$, where $a,b_\alpha$ are the
parameters of the transformation. The corresponding five conserved Noether currents are given by
\begin{equation}
j^\mu(x)=\partial^\mu\phi(x),\qquad
j^\mu_\alpha(x)=x_\alpha\partial^\mu\phi(x)-\delta^\mu_\alpha\phi(x).
\label{nonunicharges}
\end{equation}
Thanks to the commutation relations with the field operator, $[iQ,\phi(x)]=1$ and
$[iQ_\alpha,\phi(x)]=x_\alpha$, all five integral charges $Q,Q_\alpha$ are spontaneously broken,
while there is obviously only one massless mode in the spectrum. The resolution of this apparent
paradox is that the charge densities $j^0_\alpha(x)$ do not satisfy Eq.~\eqref{translation}. In fact,
a short explicit calculation gives $j^\mu_\alpha(x)-e^{iP\cdot x}j^\mu_\alpha(0)e^{-iP\cdot x}=
x_\alpha\partial^\mu\phi(x)$. Thus the standard proof of the Goldstone theorem 
does not apply to the charges $Q_\alpha$, and we should
not \emph{a priori} expect additional NG bosons stemming from their spontaneous breaking. 

An intuitive understanding of why there is only one NG boson instead of the naively expected five,
can be gained using the argument of Low and Manohar~\cite{Low:2001bw}. Since the NG boson can be
generated by a \emph{local} broken symmetry transformation acting on the order parameter for SSB,
two spontaneously broken transformations will give rise to only one NG boson if their local forms
coincide. This is exactly the case of the transformations generated by $Q,Q_\alpha$ which all have
the same local form. A similar example is a crystalline solid where both continuous translations and
rotations are spontaneously broken, yet only NG bosons corresponding to the translations---the
phonons---appear in the spectrum.

Despite the above example of an internal nonuniform symmetry, typical representatives of the
class of nonuniform symmetries according to our definition are spacetime symmetries such as
rotational or conformal invariance. In fact, the only example of a uniform spacetime symmetry is
obviously translational invariance. In the following, we will assume that there is at least a
discrete unbroken translational invariance, such as in crystalline solids. This is necessary in
order to have quasiparticles with well-defined (real or crystal) momentum. With the above in mind, our result
will apply to all spontaneously broken global continuous uniform symmetries. Eq.~\eqref{translation} 
then holds whether continuous translational invariance is spontaneously broken or not.

One might think that the restriction to uniform symmetries could be avoided by resorting to the variant of the proof of the Goldstone theorem using the quantum effective potential or action~\cite{Goldstone:1962es}, which does not rely on the operator identity \eqref{translation}. As it may be of general interest, we provide in Appendix~\ref{app:nonuniform} a modification of this proof that applies to nonuniform symmetries, leading essentially to the same conclusion as the classical argument of Ref.~\cite{Low:2001bw}. However, this method of proof only gives the number of flat directions of the effective potential (or equivalently the number of zero modes of the inverse propagator of the theory at zero momentum), which in Lorentz-noninvariant theories is in general not equal to the number of NG bosons. This means that while it can be used to make statements within Lorentz-invariant theories, it can tell us very little about the number of NG bosons in Lorentz-noninvariant theories.


\subsection{Conserved charges and their densities}

Here we make a number of other short comments related to the conserved charges required by the
Goldstone theorem. First, despite the fact that one (including ourselves) usually speaks of ``broken
generators'', it should be noted that the operators $Q_a$ in Eq.~\eqref{M} need not be generators of
symmetry transformations in the Noether sense \footnote{To avoid misunderstanding, let us emphasize
that every conserved charge of course generates a symmetry transformation on the phase space of the
classical theory. Nevertheless, it may not be possible to obtain this charge as a consequence of an
\emph{off-shell} invariance of the action in the Lagrangian formalism. This is what we mean by
saying that the charge is not of the Noether type.}. In fact, the only two properties needed for the
proof of the Goldstone theorem are that $Q_a$ are integrals or sums of charge densities as in
Eq.~\eqref{Qdef} and that $Q_a$ are time-independent. Of course, the most convenient way to ensure
that $Q_a$ is time-independent is starting from a four-current $j^\mu_a(x)$ which satisfies the
continuity equation, $\partial_\mu j^\mu_a=0$.

Second, our conjecture \eqref{conjecture} is formulated in terms of a commutator of charges.
Since generators of symmetry transformations span a Lie algebra, this strongly suggests that
$\rho_{ab}$ is actually a linear combination of expectation values of the charges themselves.
However, this would disregard the possibility that the representation of the Lie algebra of
conserved charges on the Hilbert space possesses a set of \emph{central charges}. One can then
expect a generalized commutation relation of the type $[Q_a,Q_b]=ic_{ab}\openone+if_{abc}Q_c$.
This phenomenon is certainly rather rare since the central charges $c_{ab}$ can be removed by a
suitable redefinition of the generators $Q_a$ for all semisimple Lie
algebras~\cite{Weinberg:1995v1}. Yet, once central charges do appear, it can even happen that
$\rho_{ab}$ is nonzero although at the classical level the Lie algebra of conserved charges is
commutative. The simplest example is perhaps the free non-relativistic particle which can be
interpreted as a type-II NG boson of an extended $\gr{ISO}(2)$ symmetry~\cite{Brauner:2010wm}.

Finally, it may be instructive to emphasize that any proof of the Goldstone theorem or its
generalizations, including ours, can be applied not only to the set of all broken generators of a
given theory, but also to its subsets closed under commutation. Any conclusion about the number of
NG bosons then refers to those NG bosons that couple to the broken charges considered.


\section{Charge commutator and the number of NG bosons}
\label{sec:proof}

The aim of this section is to provide evidence for our conjecture \eqref{conjecture}. We will
actually prove a one-sided inequality and give a physical argument why the opposite inequality also
holds. We will essentially follow the strategy of Nielsen and Chadha~\cite{Nielsen:1975hm} and
complement it with the relation to the charge commutator matrix $\rho_{ab}$.

The NG mode with zero momentum corresponding to the broken generator $Q_a$ will be represented as $\ket a\equiv Q_a\ket0$. This is a consistent definition of zero-momentum NG states. On the one hand, all states $\ket a$ are created by the broken charges and have zero energy. On the other hand, assume that there is another zero-energy state which is orthogonal to all $\ket a$'s. By construction, it then has a zero coupling to all broken charges, hence does not contribute to the commutator~\eqref{M}, and thus should not be counted among the NG bosons. Consequently, the number of NG bosons is equal to the number of linearly independent vectors $\ket a$. Note that this argument uses the implicit assumption that the NG states form a vector space, which is rigorously justified at least at zero momentum.


\subsection{Proof of $\nbs-\nng\leq(1/2)\rank\rho$}
\label{sec:inequality}

Since out of the $\nbs$ vectors $\ket a$, there are only $\nng$ linearly independent ones, there must be a set of complex coefficients $C_{pa}$, where $p=1,\dotsc,\dn$ and $\dn\equiv\nbs-\nng$, such that $C_{pa}\ket a=0$ with $\rank C=\dn$ (summations over repeated indices are implied). As pointed out by Nielsen and Chadha~\cite{Nielsen:1975hm}, the $\dn$ vectors $C_{pa}^*\ket a$ are linearly independent. Indeed, if this were not the case, there would be a set of coefficients $D_p$ such that $D_pC_{pa}^*\ket a=0$. Then, 
\begin{multline}
(D_pC_{pa}^*)M_{ab}=D_pC_{pa}^*\bra0[Q_a,\Phi_b(0)]\ket0=\\
=\bra aC_{pa}^*D_p\Phi_b(0)\ket0-\bra0\Phi_b(0)D_pC_{pa}^*\ket a=0,
\label{zeroeigen}
\end{multline}
where the matrix $M$ is defined in Eq.~\eqref{M}. Eq.~\eqref{zeroeigen} would imply that the matrix $M$ 
has a zero eigenvalue, which would contradict the assumption of SSB that $M$ is nonsingular.

We have observed that $C_{pa}Q_a\ket0=0$ for all $p=1,\dotsc,\dn$, and that
$\{C_{pa}^*Q_a|0\rangle\}_{p=1}^{\dn}$ are linearly independent. This implies that $A_p\equiv C_{pa}Q_a$
($\he A_p\equiv C_{pa}^*Q_a$) acts as the annihilation (creation) operator of a NG boson labelled by $p$.
Since $\{\he A_p|0\rangle\}_{p=1}^{\dn}$ are linearly independent, the $\dn\times\dn$ matrix $G$
defined as
\begin{equation}
G_{pq}\equiv\bra0A_p\he A_q\ket0=\bra0[A_p,\he A_q]\ket0=(Cg\he C)_{pq}
\label{innerproduct}
\end{equation}
is Hermitian and positive definite; the Hermitian matrix $g$ here is defined as $g_{ab}\equiv\braket
ab$. The matrix $G$ can be diagonalized by a unitary matrix $U$ so that
$UG\he U=\mathrm{diag}(\lambda_1,\dotsc,\lambda_{\dn})$. From Eq.~\eqref{innerproduct} we can see
that the diagonalization is achieved by the replacement $C\to UC$. Thus, we can choose $C$ from the
beginning in such a way that $G$ is already diagonal. We then have
\begin{align}
\bra0|[A_p,\he A_q]\ket0=\delta_{pq}\lambda_q\qquad\text{(no sum over $q$).}
\label{cr}
\end{align}
Note that this relation is reminiscent of a commutation relation between the annihilation and
creation operators, $[A_p,A^{\dagger}_q]\propto\delta_{pq}$.

Now we are ready to show the desired inequality. Let us define
\begin{equation}
\begin{split}
Q_p^+&=\frac12(A_p+\he A_p)=\mathrm{Re}(C_{pa})Q_a,\\
Q_p^-&=\frac1{2i}(A_p-\he A_p)=\mathrm{Im}(C_{pa})Q_a.
\end{split}
\label{qpm}
\end{equation}
These make $2\dn$ independent broken generators. In order to obtain a basis of the space spanned by
the broken generators $\{Q_a\}_{a=1}^{\nbs}$, we need to add $\nbs-2\dn$ other broken generators,
$Q_s'=B_{sa}Q_a$, with real $B_{sa}$. Introducing finally the notation for the new basis
\begin{multline}
(\bar{Q}_1,\bar{Q}_2,\dotsc,\bar{Q}_{\nbs})\equiv\\
\equiv(Q_1^+,Q_1^-,\dotsc,Q_{\dn}^+,Q_{\dn}^-,Q_1',\dotsc,Q_{\nbs-2\dn}'),
\label{auxeq}
\end{multline}
we infer that the antisymmetric matrix $\bar\rho$, defined by
$i\bar\rho_{ab}\equiv\lim_{\Omega\to\infty}\frac1\Omega\bra0[\bar Q_a,\bar Q_b]\ket0$, equals
\begin{equation}
\bar{\rho}=\lim_{\Omega\to\infty}\frac{1}{2\Omega}
\left(
\begin{array}{cccccc|ccc}
0&\lambda_1&&&&&&&\\
-\lambda_1&0&&&0&&&&\\
&&&\,\,\ddots&&&&{\ast}&\\
&0&&&0&\lambda_{\dn}&&&\\
&&&&-\lambda_{\dn}&0&&&\\\hline
&&&&&&&&\\
&&&{\ast}&&&&{\ast}&
\end{array}
\right),
\label{matrix}
\end{equation}
where the upper left block is $2\dn\times2\dn$ and the asterisks indicate unknown, possibly nonzero
entries. Since the transformation $\{\bar{Q}_a\}_{a=1}^{\nbs}\leftrightarrow \{Q_a\}_{a=1}^{\nbs}$
is nothing but a change of basis, $\bar\rho$ must have the same rank as $\rho$, hence
\begin{align}
\rank\rho=\rank\bar\rho\geq2\dn,
\label{final1}
\end{align}
which was to be proven.


\subsection{Explanation of $\nbs-\nng=(1/2)\rank\rho$ and the relation to the NC classification}
\label{sec:explanation}

In Ref.~\cite{Nambu:2004na}, Nambu pointed out that $\bra0[Q_a, Q_b]\ket0\neq 0$ implies that the
corresponding zero modes behave like canonical conjugates of each other; the same observation was
made in a special case in Ref.~\cite{Schafer:2001bq}. We have seen this above; $Q_p^+$ and $Q_p^-$
excite the same mode $\he A_p\ket0$, and they have the commutation relation
$\bra0[Q_p^+,Q_p^-]\ket0=(i/2)\bra0[A_p,\he A_p]\ket0\neq0$, see Eqs.~\eqref{cr} and \eqref{qpm}.

By construction, $\{Q_p^+\}_{p=1}^{\dn}\cup\{Q_s'\}_{s=1}^{\nbs-2\dn}$ (or equivalently
$\{Q_p^-\}_{p=1}^{\dn}\cup\{Q_s'\}_{s=1}^{\nbs-2\dn}$) excite linearly independent modes. So, if we
assume that Nambu's argument is valid, all of $\bra0[Q_p^+,Q_s']\ket0$, $\bra0[Q_p^-,Q_s']\ket0$ and
$\bra0[Q_{s}',Q_{s'}']\ket0$ should vanish, which means all the asterisks in Eq.~\eqref{matrix}
should be replaced with zero block matrices. We then arrive at the equality
$\nbs-\nng=(1/2)\rank\rho$.

As for the dispersion relation $\varepsilon_p(\bm{k})$ of the mode $A_p^{\dagger}\ket0$, we can
apply the same analyticity argument as Nielsen and Chadha~\cite{Nielsen:1975hm} and conclude that
$\varepsilon_p(\vek k)$ is an analytic function of $\vek k$. In the special case of an unbroken rotational invariance, 
we then have $\varepsilon_p(\vek k)\propto |\vek k|^{2m_p}$ where $m_p$ is a positive integer. Thus, all
$A_p^{\dagger}\ket0$ modes are type-II NG bosons. How about the modes excited by
$\{Q_s'\}_{s=1}^{\nbs-2\dn}$? Based on the examples discussed in literature, we can say that they
usually have a linear dispersion relation, that is, are type-I. However, there is no general
argument that would guarantee this. It is possible to modify the dispersion relation to a quadratic
one by fine tuning the parameters of the theory. Indeed, this is demonstrated by the example we
propose in Appendix~\ref{app:inequality}. All in all, the number of NG bosons with an even power of
momentum in the dispersion relation is at least $\dn$. This leads immediately to the NC inequality.

In practice, the classification of the various types of NG bosons can be significantly simplified by choosing an appropriate basis of the broken generators. To that end, note that \emph{one can always choose the ground state in such a way that only mutually commuting charges have nonzero vacuum expectation value}. In other words, only generators from the Cartan subalgebra of the Lie algebra of the symmetry group can have a nonzero vacuum expectation value. A detailed proof of this statement is given in Appendix~\ref{app:charges} at the level of generality sufficient for all practical purposes. By means of the general root decomposition of Lie algebras, it then follows immediately that the set of broken generators splits into pairs whose commutator has possibly nonzero vacuum expectation value (see Appendix~\ref{app:alternative} for details).

A cautious reader may have noticed that, strictly speaking, the above choice of basis may not be compatible with the choice made in Eq.~\eqref{auxeq}. Therefore, the material of Appendix~\ref{app:charges} cannot be directly used in our proof. However, as already stressed, it is of great practical utility and in addition, it can be used for an alternative derivation of the inequality $\nbs-\nng\leq(1/2)\rank\rho$, which is given in Appendix~\ref{app:alternative}. This derivation provides us with a new insight in the nature of the NG bosons associated with charge density.


\section{Some speculations and outlook}
\label{sec:conclusions}

In this paper we pointed out the intimate connection between the number of NG bosons in a quantum many-body system exhibiting SSB and the densities of conserved charges in the ground state. Our main result is summarized in Eq.~\eqref{conjecture}. The method used to (partially) prove it, following closely Ref.~\cite{Nielsen:1975hm}, suggests that a different classification of NG bosons than that based on the dispersion relation, might be useful. Let us denote a NG boson that couples to some combination of broken charges $C_{pa}^*Q_a$ introduced in Section~\ref{sec:proof} as type-C (``charged''), and the remaining NG bosons as type-N (``neutral''). The argument in Section~\ref{sec:proof} then immediately proves that 
\begin{equation}
n_{\mathrm N}+2n_{\mathrm C}=\nbs,
\end{equation}
with obvious notation. Moreover, our conjecture~\eqref{conjecture} can then be cast in the form $n_{\mathrm C}=(1/2)\rank\rho$.

As follows from the proof of the Nielsen--Chadha theorem~\cite{Nielsen:1975hm}, all type-C NG bosons are necessarily type-II. On the other hand, as pointed out already in Section~\ref{sec:explanation}, type-N NG bosons can be both type-I and type-II. In the systems usually discussed in literature, all type-N NG bosons actually are type-I, hence the two classifications coincide and the NC inequality is saturated as a consequence of Eq.~\eqref{conjecture}. However, the explicit example in Appendix~\ref{app:inequality} shows that there may exist NG bosons which are simultaneously type-N and type-II. We may say that such NG bosons are ``accidentally'' type-II since their nonlinear, typically quadratic, dispersion relation is achieved by tuning the parameters of the model. On the contrary, all type-C NG bosons are ``robustly'' type-II due to the connection with the conserved charge densities. The physical origin of these two different kinds of type-II NG bosons is obviously very different.

Of course, in order that this classification of NG bosons is consistently defined, one first has to
prove that it does not depend on the choice of the operators $Q_a$. For instance, the type-II NG
boson of Appendix~\ref{app:inequality} would be classified as type-N. However, one could imagine
that there is another conserved charge which couples to the the same NG boson and has a nonzero
commutator with the Noether charge of the shift symmetry of the theory \eqref{laginequ}. The same NG
boson would then be classified as type-C. Of course, the conjecture \eqref{conjecture} would be
satisfied in both cases. That is why we only propose this classification here in the form of a
speculation. It would be interesting to investigate this issue further.

Another interesting problem is what happens when symmetry whose spontaneous breaking gives rise to
a reduced number of NG bosons is gauged. How many massive spin-one bosons appear in the spectrum as
a consequence of the Anderson--Higgs mechanism? Is their number equal to $\nng$ or rather to $\nbs$?
This question has been addressed in
Refs.~\cite{Gusynin:2003yu,Hama:2011rt}, but so far only for one particular model, whose global
version was introduced earlier in Refs.~\cite{Schafer:2001bq,Miransky:2001tw}. The problem would
certainly deserve a more general investigation.

Finally, note that in this paper we concentrated solely on NG bosons of spontaneously broken 
uniform symmetries. While a generalization of the Goldstone theorem to spontaneously broken
nonuniform symmetries, which ensures the existence of at least one NG boson, was given in
Appendix~\ref{app:nonuniform}, the issue of the count of the NG bosons in such a case is still
open. Any result with the same level of generality as well as rigor as the Goldstone theorem
would significantly improve our understanding of spacetime as well as other nonuniform symmetries.


\begin{acknowledgments}
We thank H.~Abuki, H.~Aoki, T.~Hatsuda, T.~Hayata, T.~Kanazawa, and T.~Kugo for fruitful discussions. Part of the work
was carried out during the stay of H.W.~at the University of Bielefeld. H.W.~acknowledges the University 
of Tokyo exchange program for a partial support of his visit to Bielefeld. The work
of T.B.~was supported by the Sofja Kovalevskaja program of the Alexander von Humboldt Foundation. 
\end{acknowledgments}

\appendix


\section{Inequality in the NC counting rule}
\label{app:inequality}

Here we propose an example in which a strict inequality in the NC counting rule holds.
Similarly to the example in Section~\ref{sec:uniform}, let us consider a class of free scalar field
theories defined by the Lagrangian density
\begin{equation}
\La=\frac{1}{2}\partial_0\phi\partial_0\phi-\frac{1}{2}\phi\mathcal{D}\phi,
\label{laginequ}
\end{equation}
where $\mathcal{D}=\sum_{n=1}^{\infty}c_n(-\vek\nabla^2)^n$. This Lagrangian is invariant under
a (real) constant shift of the field, $\phi(x)\rightarrow \phi(x)+\theta$.

By choosing the ground state $\ket0$ for simplicity as the Fock vacuum, we get 
\begin{equation}
\bra0|[iQ, \phi(x)]\ket0=1.
\end{equation}
Thus, the shift symmetry is \emph{always} spontaneously broken. The corresponding NG boson is
nothing but the one-particle state in the Fock space of this noninteracting field theory, 
$\ket{n_{\vek k}}=a^{\dagger}_{\vek k}\ket0$ with the dispersion relation
$\varepsilon(\vek k)=\sqrt{\sum_{n}c_n|\vek k|^{2n}}$. For the usual Lorentz-invariant case
($c_n=\delta_{n,1}$), we have $\varepsilon(\vek k)=|\vek k|$ which represents a type-I NG boson so that
the NC inequality is saturated. However, despite being somewhat pathological, it is also possible to
choose $c_n\propto\delta_{n,2}$. In this case, $\varepsilon(\vek k)\propto|\vek k|^2$ which describes a
type-II NG boson, hence a strict inequality occurs in the NC counting rule. The punch line is that
the power of momentum in the dispersion relation of the NG boson can be modified by tuning the
parameters of the Lagrangian. We expect the same is also true in other models such as the
Heisenberg ferromagnet with a sufficient number of coupling constants.


\section{Goldstone theorem for nonuniform symmetries}
\label{app:nonuniform}

The alternative proof of the Goldstone theorem~\cite{Goldstone:1962es} usually makes use of the
quantum effective potential, which is a function of classical \emph{uniform} fields, and thus cannot
be applied directly to nonuniform symmetry transformations. We offer here a generalization of the
argument that relies on the full quantum effective action. 

Following essentially Weinberg~\cite{Weinberg:1996v2}, we consider a theory of a set of (not
necessarily scalar) fields, $\phi_i(x)$, and assume that the classical action of the theory is
invariant under the infinitesimal shift $\delta\phi_i(x)=\theta F_i[x;\phi]$. Here $\theta$ is a
parameter of the transformation, and $F_i$ is in general a \emph{functional} of the fields and a
function of the coordinate $x$. This classical symmetry is shared by the quantum effective action,
$\Gamma[\phi]$, that is,
\begin{equation}
\int d^4y\,\frac{\delta\Gamma[\phi]}{\delta\phi_j(y)}F_j[y;\phi]=0,
\label{aux1}
\end{equation}
provided $F_i$ is linear in the fields. Taking now the variational derivative of Eq.~\eqref{aux1}
with respect to $\phi_i(x)$ and setting the field equal to its expectation value,
$\bra0\phi_i(x)\ket0\equiv v_i(x)$, we obtain
\begin{equation}
\int d^4y\,\frac{\delta^2\Gamma[v]}{\delta\phi_i(x)\delta\phi_j(y)}F_j[y;v]=0.
\label{aux2}
\end{equation}
Finally, we assume for the sake of simplicity full translational invariance of the vacuum. The second derivative 
of the effective action can then be expressed in terms of the inverse propagator of the theory,
$G^{-1}_{ij}(k)$. 

For \emph{uniform} symmetries we can immediately infer that $G^{-1}_{ij}(0)F_j[v]=0$. In other
words, SSB implies the existence of a gapless pole in the propagator. This is equivalent to the more
common formulation of the proof using the effective potential. Our argument is nevertheless more
general since it also applies to theories with a nonlocal action which cannot be expressed as a
spacetime integral of a local Lagrangian density.

For nonuniform symmetries, we have to make one further step, that is, multiply Eq.~\eqref{aux2} by
$e^{ik\cdot x}$ and integrate over the coordinate $x$, which yields
\begin{equation}
G^{-1}_{ij}(k)\int d^4y\,e^{ik\cdot y}F_j[y;v]=0.
\label{manohar}
\end{equation}
(Nonzero momentum $k$ is used just to ensure the convergence of the integral.) Once again, in the
limit $k\to0$, the integral of $F_j[y;v]$ represents a zero mode of the inverse propagator, which
establishes the existence of a massless particle in the spectrum.

Applying Eq.~\eqref{manohar} to the example discussed in Section~\ref{sec:uniform}, we can
immediately see that all five spontaneously broken charges of Eq.~\eqref{nonunicharges} give rise to
the \emph{same} zero mode of the inverse propagator. Generally, it follows from Eq.~\eqref{manohar}
that two broken symmetries that have the same local form will give the same zero mode of the inverse
propagator, that is, the same NG boson. This is in accord with the conclusion of Low and
Manohar~\cite{Low:2001bw}. Nevertheless, in contrast to their classical field theory argument, our
derivation actually proves the existence of the massless mode in the spectrum of the quantum theory,
and is only limited by the assumed translational invariance and the linearity of the symmetry transformation 
in the fields.


\section{Charge densities and the Cartan subalgebra}
\label{app:charges}

In this appendix we will discuss in detail the statement made in Section~\ref{sec:explanation} that
\emph{one can always choose the ground state in such a way that only mutually commuting charges have
nonzero vacuum expectation value}. As explained there, this is of great practical help in the
classification of the NG bosons. Even though we have not been able to prove this claim in the full
generality, we give below a detailed proof that applies to all compact semisimple Lie groups which
are given by direct products of \emph{classical} simple Lie groups. This is sufficient for virtually
all practical purposes. Our claim can be reformulated mathematically as the statement that
\emph{every adjoint orbit of the symmetry group passes through the Cartan subalgebra} of its Lie
algebra. We will prove it in turn for all classes of classical simple Lie groups and address the
generalization to semisimple groups in the end.

Let us first introduce some notation. By $T_a$ and $\op Q_a$ we will denote the generators of the
symmetry group in its defining representation, and its representation on the Hilbert space of the
quantum system, respectively. The corresponding finite symmetry transformations will then be
$U(\theta)\equiv e^{i\theta_a T_a}$ and $\U(\theta)\equiv e^{i\theta_a\op Q_a}$. Finally,
$R(\theta)$ will stand for the adjoint representation of the same transformation. Note that for the
sake of clarity, we use in this appendix hats to distinguish operators on the Hilbert space from
finite-dimensional matrices.

We now introduce the charge expectation values, $q_a\equiv\bra0\op Q_a\ket0$, and construct the
matrix $\M\equiv q_aT_a$ belonging to the Lie algebra of the symmetry group. Upon a symmetry
transformation, the ground state $\ket0$ becomes $\ket\theta\equiv\U(\theta)\ket0$. The charges $q_a$
then transform to \footnote{The last step here makes use of the assumption that the adjoint
representation matrices $R(\theta)$ are orthogonal, which in turn follows from the full antisymmetry
of the structure constants of the Lie algebra. This can be achieved for any compact semisimple Lie
algebra by a proper choice of basis.}
\begin{multline}
q'_a\equiv\bra\theta\op Q_a\ket\theta=\bra0\he{\U(\theta)}\op Q_a\U(\theta)\ket0=\\
=\bra0R(\theta)^{-1}_{ab}\op Q_b\ket0=R(\theta)_{ba}q_b.
\end{multline}
Hence $\M$ transforms to
\begin{equation}
\M'\equiv q'_bT_b=q_aR(\theta)_{ab}T_b=U(\theta)\M\he{U(\theta)}.
\end{equation}
We can see that $\M$ transforms in the adjoint representation, which provides a link between the
physical and the mathematical formulation of the problem; the fact that only $q_a'$s
corresponding to mutually commuting generators are nonzero simply means that the matrix $\M'$ lies
in the Cartan subalgebra. In the following, we will use the more concise mathematical formulation of
the problem.


\subsection{Proof for the groups $\gr{SU}(N)$}

Since the Cartan subalgebra is in this case formed by all real diagonal traceless matrices, our
claim follows trivially from the well-known fact that every Hermitian matrix can be diagonalized by
a suitable (unimodular) unitary transformation. To the best of our knowledge, the proof in this case
was given for the first time in Ref.~\cite{Rajagopal:2005dg} using an argument essentially
identical to ours.


\subsection{Proof for the groups $\gr{SO}(N)$}

The Cartan subalgebra in this case can be chosen as the set of all (purely imaginary and
antisymmetric) matrices whose only nonzero entries are localized in $2\times2$ blocks along the
diagonal. Our task is to prove that every purely imaginary antisymmetric matrix $\M$ can be brought
into such block-diagonal form by a special orthogonal transformation. 

We know that the (real) eigenvalues of a purely imaginary antisymmetric matrix come in pairs with
opposite signs and that the corresponding eigenvectors are related by complex conjugation. Let us
denote these eigenvectors as $\vek u_k,\vek u^*_k$. Every such pair of vectors span an invariant
subspace of $\M$. We can trade them for real vectors $\vek v_k\equiv(\vek u_k+\vek u^*_k)/\sqrt2$
and $\vek w_k\equiv-i(\vek u_k-\vek u^*_k)/\sqrt2$. (In case of odd $N$, there will be an additional real
eigenvector associated with zero eigenvalue.) Since the basis of eigenvectors $\{\vek u_k,\vek
u^*_k\}_{k=1}^{N/2}$ is orthonormal, so is the basis $\{\vek v_k,\vek w_k\}_{k=1}^{N/2}$. It is
real and hence connected to the basis we started from by a real orthogonal transformation. In this
basis, $\M$ takes the desired block-diagonal form. Finally, by flipping the sign of one of the basis
vectors if necessary, we can ensure that the similarity transformation leading to $\{\vek v_k,\vek
w_k\}_{k=1}^{N/2}$ is unimodular.


\subsection{Proof for the groups $\gr{Sp}(2N)$}

Recall that the group $\gr{Sp}(2N)$ consists of all $2N\times2N$ unitary matrices $U$ such that $U\I
U^T=\I$ where $\I\equiv i\sigma_2\otimes\openone_{N\times N}$ with $\sigma_2$ the second Pauli
matrix. We can say that $\gr{Sp}(2N)$ contains all unitary matrices which are simultaneously
$\I$-orthogonal. Likewise, the generators of $\gr{Sp}(2N)$ are Hermitian and $\I$-antisymmetric,
$T_a^T=-\I T_a\I ^{-1} $. As a consequence, the generators of $\gr{Sp}(2N)$ take the form
\begin{equation}
T=\begin{pmatrix}
A & B\\
\he B & -A^T
\end{pmatrix},
\end{equation}
where $A$ is  a Hermitian and $B$ a symmetric complex $N\times N$ matrix. The Cartan subalgebra
then consists of all matrices of this form with $A$ real and diagonal and $B=0$.

We are now to prove that every $\I$-antisymmetric (and Hermitian) matrix $\M$ can be diagonalized
by a unitary symplectic transformation. First, observe that if $\vek u_k$ is its eigenvector with
the (real) eigenvalue $\lambda_k$, then its $\I$-conjugate, $\I\vek u_k^*$, is also an eigenvector
with the eigenvalue $-\lambda_k$. Hence the eigenvalues of $\M$ once again come in pairs with
opposite sign and we can thus construct an orthonormal basis of eigenvectors and organize them
in the columns of a unitary matrix \footnote{While the orthogonality is trivial in the case of a
nonzero eigenvalue with multiplicity one, in general one has to be a bit careful. First, note that
when $\lambda_k=0$, $\I\vek u_k^*$ is orthogonal to $\vek u_k$ by construction, thanks to the
antisymmetry of $\I$. Second, suppose that some eigenvalues are equal, say,
$\lambda_{k+1}=\lambda_k$. We can then choose the eigenvector $\vek u_{k+1}$ to be orthogonal to
both $\vek u_k$ and $\I\vek u^*_k$. This already guarantees that $\I\vek u_{k+1}^*$ will be
orthogonal to all three of these vectors. One can thus construct an orthonormal basis by induction.},
\begin{equation}
P\equiv(\vek u_1,\dotsc,\vek u_N,\I\vek u_1^*,\dotsc,\I\vek u_N^*).
\end{equation}
In this basis, $\M$ acquires the desired diagonal form and a simple calculation shows that $P$
itself is symplectic as required. This completes the proof.


\subsection{Proof for semisimple groups}

Let us prove our claim for a direct product of two groups of the type discussed in the preceding
three subsections. The general case will then follow by induction. The two (commuting) sets of
generators will be denoted $T_a\otimes\openone$ and $\openone\otimes T_\alpha$, where we use
Latin and Greek indices to distinguish the two groups. The matrix $\M$ then becomes
\begin{equation}
\M=(q_aT_a)\otimes\openone+\openone\otimes(q_\alpha T_\alpha).
\end{equation}
The transformations from the two groups are labelled by independent parameters $\theta_a$ and
$\sigma_\alpha$, that is, $U(\theta)\equiv e^{i\theta_aT_a}$ and $V(\sigma)\equiv e^{i\sigma_\alpha
T_\alpha}$. Under a general transformation from the product group, $U(\theta)\otimes V(\sigma)$,
the matrix $\M$ transforms to
\begin{multline}
\M'=[U(\theta)\otimes V(\sigma)]\M\he{[U(\theta)\otimes V(\sigma)]}=\\
=[U(\theta)q_aT_a\he{U(\theta)}]\otimes\openone+
\openone\otimes[V(\sigma)q_\alpha T_\alpha\he{V(\sigma)}].
\end{multline}
Obviously, one can choose the transformations in the two groups independently to diagonalize the
respective parts of $\M$ so that in the end, $\M'$ will lie in the Cartan subalgebra of the product
group, which is just a direct sum of the two Cartan subalgebras.


\section{Classification of NG bosons via Cartan decomposition}
\label{app:alternative}

In order to view the NG bosons associated with charge densities from a different perspective, let us recall the root decomposition from the theory of Lie algebras. (Our argument will apply to all Lie algebras to which the result of Appendix~\ref{app:charges} applies, that is, to all compact semisimple Lie algebras which are given by direct products of \emph{classical} simple Lie algebras.) There, one constructs linear combinations of the Lie algebra generators in the form of root generators $E_\alpha$, such that $[E_\alpha,\he E_\beta]=\delta_{\alpha\beta}\alpha_iH_i$, where $\alpha_i$ is the associated root vector and $H_i$ is the basis of the Cartan subalgebra.

Using the result of Appendix~\ref{app:charges}, we now choose the basis of the Lie algebra so that only the generators $H_i$ can have a nonzero vacuum expectation value. Concretely, let us consider all the roots $\alpha_c$, $c=1,\dotsc,\dn'$ such that $\lambda_c'\equiv(\alpha_c)_i\bra0H_i\ket0\neq0$. Defining analogously to Eq.~\eqref{qpm} the set of $2\dn'$ generators $Q^\pm_c$ as
\begin{equation}
\begin{split}
Q^+_c&=\frac1{2}(E_{\alpha_c}+\he E_{\alpha_c}),\\
Q^-_c&=\frac1{2i}(E_{\alpha_c}-\he E_{\alpha_c}),
\end{split}
\end{equation}
and adding $\nbs-2\dn'$ generators $Q_s'$ in order to obtain a complete basis of generators, the commutator matrix $\rho$ becomes
\begin{equation}
\rho=\lim_{\Omega\to\infty}\frac{1}{2\Omega}
\left(
\begin{array}{cccccc|ccc}
0&\lambda'_1&&&&&&&\\
-\lambda'_1&0&&&0&&&&\\
&&&\,\,\ddots&&&&0&\\
&0&&&0&\lambda'_{\dn'}&&&\\
&&&&-\lambda'_{\dn'}&0&&&\\\hline
&&&&&&&&\\
&&&0&&&&0&
\end{array}
\right).
\end{equation}
Ordering the generators as in Eq.~\eqref{auxeq}, we next choose the first $2\dn'$ interpolating fields $\Phi_a$ in the commutator~\eqref{M} as $(j^{0+}_1,j^{0-}_1,\dotsc,j^{0+}_{\dn'},j^{0-}_{\dn'})$. For $s=2\dn'+1,\dotsc,\nbs$ and $a=1,\dotsc,2\dn'$, we now have $M_{sa}=0$, and the SSB condition $\det M\neq0$ demands that $\det M'\neq0$ where $M'_{st}$ is the $(\nbs-2\dn')\times(\nbs-2\dn')$ lower-right corner of $M_{ab}$. Finally, we observe that since the expectation values of commutators of all pairs of generators $Q_s'$ vanish, according to the theorem of Sch\"afer~\emph{et al.}~\cite{Schafer:2001bq} there must be $\nbs-2\dn'$ NG bosons as intermediate states in the commutator $M'_{st}$. At the same time, at least one NG boson must appear as an intermediate state in the commutator $\bra0[Q^+_c,j^{0-}_c(0)]\ket0$ for each $c=1,\dotsc,\dn'$. Therefore, the number of NG bosons as bounded from below as $\nng\geq(\nbs-2\dn')+\dn'=\nbs-\dn'=\nbs-(1/2)\rank\rho$, as proved by a different method in Section~\ref{sec:inequality}. While the above argument uses the notation for charge densities introduced in Section~\ref{sec:generalSSB} for systems with continuous translational invariance, it goes through without change if we replace $j^0_a$ with $\rho_a$---a charge density on a spatial lattice---as in Eq.~\eqref{Qdef}.


\bibliography{references}

\end{document}